\documentclass[12pt]{article}
\usepackage{axodraw,bbold}

\catcode`@=12
\begin{document}
\thispagestyle{empty}
\begin{flushright} 
UCRHEP-T444\\ 
January 2008\
\end{flushright}
\vspace{0.5in}
\begin{center}
{\LARGE	\bf Supersymmetric U(1) Gauge Realization of\\ the Dark Scalar Doublet
Model of\\ Radiative Neutrino Mass\\}
\vspace{1.0in}
{\bf Ernest Ma\\}
\vspace{0.2in}
{\sl Department of Physics and Astronomy, University of California,\\}
\vspace{0.1in}
{\sl Riverside, California 92521, USA\\}
\vspace{1.0in}
\end{center}

\begin{abstract}\
Adding a second scalar doublet $(\eta^+,\eta^0)$ and three neutral singlet 
fermions $N_{1,2,3}$ to the Standard Model of particle interactions with a 
new $Z_2$ symmetry, it has been shown that $\eta^0_R$ or $\eta^0_I$ is a 
good dark-matter candidate and seesaw neutrino masses are generated 
radiatively.  A supersymmetric U(1) gauge extension of this new idea is 
proposed, which enforces the usual $R$ parity of the Minimal Supersymmetric 
Standard Model, \emph{and} allows this new $Z_2$ symmetry to emerge as a 
discrete remnant.
\end{abstract}

\newpage
\baselineskip 24pt

A new idea has recently been proposed \cite{m06-1,m06-2,m06-3} that without 
dark matter, neutrinos would be massless.  This is minimally implemented in 
the Standard Model (SM) of particle interactions by the addition of a second 
scalar doublet $(\eta^+,\eta^0)$ and three neutral singlet Majorana fermions 
$N_{1,2,3}$, together with a new $Z_2$ discrete symmetry 
\cite{dm78}, under which $(\eta^+,\eta^0)$ and $N_{1,2,3}$ are odd, and all 
SM particles are even.  Using the allowed term 
$(\lambda_5/2)(\Phi^\dagger \eta)^2 + H.c.$, where $\Phi=(\phi^+,\phi^0)$ is 
the SM Higgs doublet, seesaw neutrino masses are 
generated in one loop, as shown in Fig.~1.

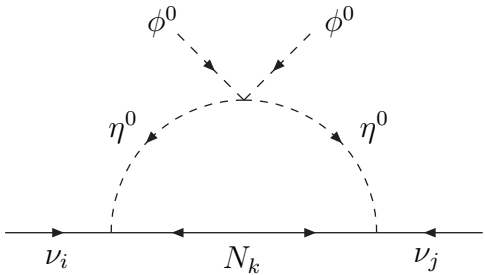
\begin{figure}[htb]
\begin{center}
\begin{picture}(360,120)(0,0)
\ArrowLine(90,10)(130,10)
\ArrowLine(180,10)(130,10)
\ArrowLine(180,10)(230,10)
\ArrowLine(270,10)(230,10)
\DashArrowLine(155,85)(180,60)3
\DashArrowLine(205,85)(180,60)3
\DashArrowArc(180,10)(50,90,180)3
\DashArrowArcn(180,10)(50,90,0)3

\Text(110,0)[]{$\nu_i$}
\Text(250,0)[]{$\nu_j$}
\Text(180,0)[]{$N_k$}
\Text(135,50)[]{$\eta^0$}
\Text(230,50)[]{$\eta^0$}
\Text(150,90)[]{$\phi^{0}$}
\Text(217,90)[]{$\phi^{0}$}

\end{picture}
\end{center}
\caption{One-loop generation of neutrino mass.}
\end{figure}

\noindent At the same time, $\eta^0_R$ and $\eta^0_I$ are split in mass, and 
whichever is lighter has been shown to be a good dark-matter candidate 
\cite{bhr06,lnot07,glbe07} with mass between 45 and 75 GeV, with reasonable 
prognosis \cite{bhr06,cmr07} for detection at the Large Hadron Collider (LHC). 
Variants of this basic idea have also been discussed 
\cite{knt03,cs04,kms06,ks06,hkmr07,ss07,bm07,m07-1,m07-2}.

Consider now its supersymmetric generalization.  The simplest version 
\cite{m06-3,m07-3} is to retain the proposed $Z_2$ symmetry in addition to the 
usual $R$ parity of the Minimal Supersymmetric Standard Model (MSSM) which 
forbids proton decay.  Another is to use the string-inspired $E_6/U(1)_N$ model 
\cite{ms07} but it is still necessary to impose the two $Z_2$ symmetries 
by hand.  In either case, there are two kinds of dark matter as recently 
discussed \cite{hln07,cmwy07}.

In this paper, instead of imposing the two $Z_2$ symmetries by hand, a new 
U(1) gauge symmetry is proposed, from which $R$ parity is automatically 
derived.  Further, as this U(1) gauge symmetry is spontaneously broken, its 
residual $Z_2$ discrete symmetry \cite{ks06} is exactly what is needed for 
radiative neutrino mass and the second type of dark matter.

Consider the following U(1) gauge extension of the MSSM with particle content 
as shown in Table 1.

\begin{table}[htb]
\caption{MSSM particle content of proposed model.}
\begin{center}
\begin{tabular}{|c|c|c|}
\hline 
Superfield & $SU(3)_C \times SU(2)_L \times U(1)_Y$ & $U(1)_X$ \\ 
\hline
$Q \equiv (u,d)$ & $(3,2,1/6)$ & $n_1$ \\ 
$u^c$ & $(3^*,1,-2/3)$ & $n_2$ \\ 
$d^c$ & $(3^*,1,1/3)$ & $n_3$ \\ 
$L \equiv (\nu,e)$ & $(1,2,-1/2)$ & $n_4$ \\ 
$e^c$ & $(1,1,1)$ & $n_1+n_3-n_4$ \\ 
\hline
$\Phi_1 \equiv (\phi^0_1,\phi^-_1)$ & $(1,2,-1/2)$ & $-n_1-n_3$ \\ 
$\Phi_2 \equiv (\phi^+_2,\phi^0_2)$ & $(1,2,1/2)$ & $-n_1-n_2$ \\ 
\hline
\end{tabular}
\end{center}
\end{table}

Without $U(1)_X$, the following terms are allowed:
\begin{equation}
L\Phi_2, ~~ LLe^c, ~~ LQd^c, ~~ u^cd^cd^c,
\end{equation}
thereby violating both lepton number ($L$) and baryon number ($B$).  In the 
MSSM, they are eliminated by the imposition of $R$ parity, i.e. 
$R \equiv (-)^{3B+L+2j}$.  Suppose $U(1)_X$ is used instead, then
\begin{equation}
n_1=n_4=0, ~~~ n_2=-n_3
\end{equation}
forbids these terms, and satisfies all the anomaly-free conditions for 
$U(1)_X$ to be a gauge symmetry except one, i.e. $[U(1)_X]^3$, which reads
\begin{equation}
3[6 n_1^3 + 3 n_2^3 + 3 n_3^3 + 2 n_4^3 + (n_1+n_3-n_4)^3] + 2 (-n_1-n_3)^3 
+ 2 (-n_1-n_2)^3 = -3 n_2^3.
\end{equation}
To cancel this contribution, the usual solution is to introduce 3 neutral 
singlet fermions with $n_2$ as their $U(1)_X$ assignment, i.e. the canonical 
minimal extension with three right-handed singlet neutrinos where $U(1)_X$ is 
identified as $T_{3R}$ and $n_2 = -1/2$.  Here a different solution is 
proposed so that the dark scalar doublet model of radiative neutrino mass 
\cite{m06-1,m06-2,m06-3} may be realized.  Two doublet superfields and 
several singlet superfields are added, as shown in Table 2.

\begin{table}[htb]
\caption{New particle content of proposed model.}
\begin{center}
\begin{tabular}{|c|c|c|}
\hline 
Superfield & $SU(3)_C \times SU(2)_L \times U(1)_Y$ & $U(1)_X$ \\ 
\hline
$\eta_1 \equiv (\eta^0_1,\eta^-_1)$ & $(1,2,-1/2)$ & $n_2/4$ \\ 
$\eta_2 \equiv (\eta^+_2,\eta^0_2)$ & $(1,2,1/2)$ & $-n_2/4$ \\ 
\hline
$N_i$ & $(1,1,0)$ & $n_2/4$ \\
$\chi$ & $(1,1,0)$ & $-3n_2/4$ \\
\hline
$S$ & $(1,1,0)$ & $-n_2/2$ \\
$\zeta$ & $(1,1,0)$ & $3n_2/2$ \\
\hline
\end{tabular}
\end{center}
\end{table}

Consequently, the $[U(1)_X]^3$ anomaly is given by
\begin{equation}
3(-1)^3 + 2 \left( {1 \over 4} \right)^3 + 2 \left( -{1 \over 4} \right)^3 
+ n_N \left( {1 \over 4} \right)^3 + \left( -{3 \over 4} \right)^3 + 
\left( -{1 \over 2} \right)^3 + \left( {3 \over 2} \right)^3 = (n_N-11) 
\left( {1 \over 4} \right)^3,
\end{equation}
together with the mixed gravitational-gauge $U(1)_X$ anomaly
\begin{equation}
3(-1) + 2 \left( {1 \over 4} \right) + 2 \left( -{1 \over 4} \right) 
+ n_N \left( {1 \over 4} \right) + \left( -{3 \over 4} \right) + 
\left( -{1 \over 2} \right) + \left( {3 \over 2} \right) = (n_N-11) 
\left( {1 \over 4} \right),
\end{equation}
where $n_N$ is the number of $N$ superfields.  It is clear that $n_N=11$ 
is the simplest solution for an anomaly-free $U(1)_X$ in this case. 
From the assignments of Table 2, there are now the following 
allowed terms:
\begin{eqnarray}
L \eta_2 N, ~~ \Phi_1 \eta_2 \chi, ~~ \eta_1 \eta_2, ~~ N N S, 
~~ \chi \chi \zeta.
\end{eqnarray}

The one-loop radiative generation of neutrino mass is then possible, as shown 
in Fig.~2.  Assuming that $U(1)_X$ is broken spontaneously by the nonzero 
vacuum expectation values of $S$ and $\zeta$, a discrete $Z_2$ symmetry 
remains, under with $N,\chi,\eta$ are odd, and all other superfields are even. 
This allows a linear combination of $\eta^0_{1,2}$ and $\chi$ 
(call it $\chi_0$) to be 
a dark-matter candidate in addition to the lightest neutralino. As for the 
usual $R$ parity, it remains in effect by assigning $N$ to be odd under 
$(-)^L$ and all other new superfields to be even.  Note also that because 
there are only three families of $L$, only three out of the eleven $N$'s 
couple to $L$ through $\eta_2$.

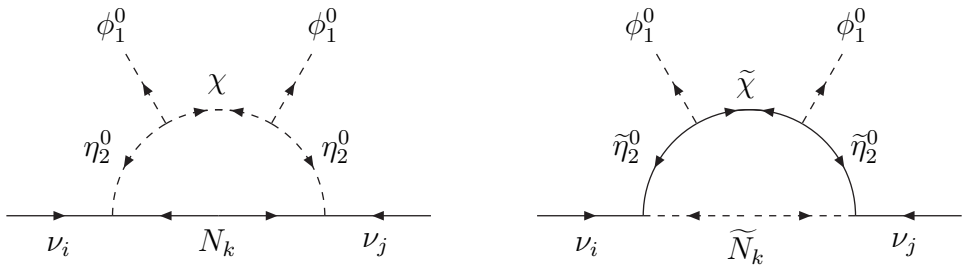
\begin{figure}[htb]
\begin{center}\begin{picture}(500,100)(10,45)
\ArrowLine(70,50)(110,50)
\ArrowLine(150,50)(190,50)
\ArrowLine(150,50)(110,50)
\ArrowLine(230,50)(190,50)
\Text(90,35)[b]{$\nu_i$}
\Text(210,35)[b]{$\nu_j$}
\Text(150,35)[b]{$N_k$}
\Text(150,100)[]{$\chi$}
\Text(105,70)[b]{$\eta^0_2$}
\Text(195,70)[b]{$\eta^0_2$}
\Text(110,116)[b]{$\phi^0_1$}
\Text(190,116)[b]{$\phi^0_1$}
\DashArrowLine(130,85)(115,111){3}
\DashArrowLine(170,85)(185,111){3}
\DashArrowArc(150,50)(40,120,180){3}
\DashArrowArc(150,50)(40,60,100){3}
\DashArrowArcn(150,50)(40,60,0){3}
\DashArrowArcn(150,50)(40,120,80){3}

\ArrowLine(270,50)(310,50)
\DashArrowLine(352,50)(390,50){3}
\DashArrowLine(348,50)(310,50){3}
\ArrowLine(430,50)(390,50)
\Text(290,35)[b]{$\nu_i$}
\Text(410,35)[b]{$\nu_j$}
\Text(350,32)[b]{$\widetilde N_k$}
\Text(350,100)[]{$\widetilde \chi$}
\Text(305,70)[b]{$\widetilde \eta^0_2$}
\Text(395,70)[b]{$\widetilde \eta^0_2$}
\Text(310,116)[b]{$\phi^0_1$}
\Text(390,116)[b]{$\phi^0_1$}
\DashArrowLine(330,85)(315,111){3}
\DashArrowLine(370,85)(385,111){3}
\ArrowArc(350,50)(40,120,180)
\ArrowArc(350,50)(40,60,100)
\ArrowArcn(350,50)(40,60,0)
\ArrowArcn(350,50)(40,120,80)

\end{picture}
\end{center}
\caption[]{One-loop radiative contributions to neutrino mass.}
\end{figure}

Because of the two separately conserved discrete symmetries, there are now 
at least two absolutely stable particles \cite{m06-3,m07-3,ms07,hln07,cmwy07}: 
the lightest particle with $R=-1$ as in the MSSM, and the lightest particle 
which is odd under the $Z_2$ remnant of $U(1)_X$. Consider in particular the 
three lightest particles with $(R,Z_2) = (-,+)$, $(+,-)$, and $(-,-)$ 
respectively. If one is heavier than the other two combined, then the latter 
are the two components of dark matter.  If not, then all three contribute. 

With the particle content of Table 2, the question of fermion masses requires 
some discussion.  It is clear that $\eta_1 \eta_2$ is an allowed mass term, 
whereas $N$ and $\chi$ obtain Majorana masses through the $N N S$ and 
$\chi \chi \zeta$ couplings, as ${S}$ and ${\zeta}$ acquire nonzero vacuum 
expectation values.  As for $\widetilde{S}$ and $\widetilde{\zeta}$ 
themselves, a Dirac mass term links the linear combination $(3v_\zeta 
\widetilde{\zeta} - v_S \widetilde{S})/\sqrt{9v_\zeta^2+v_S^2}$ to the $U(1)_X$ 
gaugino, whereas $(3 v_\zeta \widetilde{S} + v_S \widetilde{\zeta})/
\sqrt{9v_\zeta^2+v_S^2}$ (call it $\widetilde{S_0}$) remains massless at 
tree level.  To remedy this situation, one way is to add two singlet 
superfields $\delta_{1,2} \sim \pm 3n_2$, then $\delta_1 \delta_2$ and 
$\zeta \zeta \delta_2$ are allowed terms, and there will be no massless 
particle at tree level.  On the other hand, even without adding anything, 
$\widetilde{S}$ and $\widetilde{\zeta}$ have radiative mass terms in one 
loop as shown in Fig.~3.
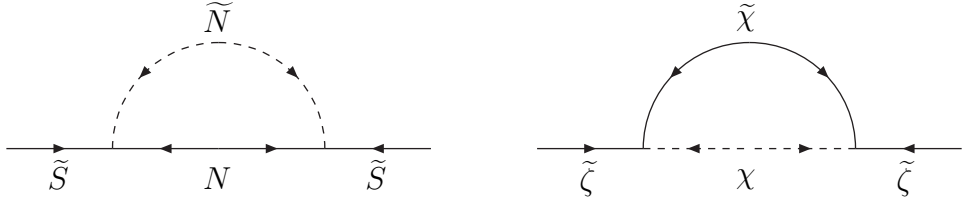
\begin{figure}[htb]
\begin{center}
\begin{picture}(500,80)(10,45)
\ArrowLine(70,50)(110,50)
\ArrowLine(150,50)(190,50)
\ArrowLine(150,50)(110,50)
\ArrowLine(230,50)(190,50)
\Text(90,35)[b]{$\widetilde{S}$}
\Text(210,35)[b]{$\widetilde{S}$}
\Text(150,35)[b]{$N$}
\Text(150,100)[]{$\widetilde{N}$}
\DashArrowArc(150,50)(40,90,180){3}
\DashArrowArcn(150,50)(40,90,0){3}

\ArrowLine(270,50)(310,50)
\DashArrowLine(352,50)(390,50){3}
\DashArrowLine(348,50)(310,50){3}
\ArrowLine(430,50)(390,50)
\Text(290,32)[b]{$\widetilde{\zeta}$}
\Text(410,32)[b]{$\widetilde{\zeta}$}
\Text(350,35)[b]{$\chi$}
\Text(350,100)[]{$\widetilde \chi$}
\ArrowArc(350,50)(40,90,180)
\ArrowArcn(350,50)(40,90,0)
\end{picture}
\end{center}
\caption[]{One-loop radiative contributions to $S$ and $\zeta$ Majorana 
masses.}
\end{figure}
The $4 \times 4$ neutralino mass matrix of the MSSM is then extended to a 
$7 \times 7$ matrix, with the inclusion of the $U(1)_X$ gaugino as well as 
the $\widetilde{S}$ and $\widetilde{\zeta}$ higgsinos, and $\widetilde{S_0}$ 
is expected to be the lightest neutralino and the dark-matter candidate 
with $(R,Z_2) = (-,+)$.  However, it interacts with quarks through the 
$U(1)_X$ gauge boson with the effective interaction $6(v_\zeta - v_S) 
(9 v_\zeta^2 + v_S^2)^{-3/2}$ which has to be suppressed by a factor of 
about $10^{-4}$ relative to $G_F$ for it to satisfy the present upper bound 
on the direct detection of dark matter in underground laboratory 
experiments.  This may be accomplished for example with $v_\zeta \sim v_S \sim 
10~G_F^{-1/2}$ and $v_\zeta - v_S \sim G_F^{-1/2}$.  Since there are eleven 
$N$'s, the radiative mass of $\widetilde{S}$ in Fig.~3 could be enhanced by 
an order of magnitude, and for $v_S \sim 1$ TeV, $m_{\widetilde{S_0}}$ may be 
as high as 100 GeV.  In that case, $\widetilde{S_0} \widetilde{S_0}$ may 
annihilate to $\chi_0 \chi_0$ through the $\chi \chi \zeta$ coupling more 
efficiently than through its supressed coupling to the $U(1)_X$ gauge boson 
itself.

Since the $U(1)_X$ gauge boson $X$ interacts with both quarks and leptons, 
its production is similar to that of other U(1) gauge bosons, such as those 
from $E_6$.  As already mentioned, as far as the quarks and charged leptons 
are concerned, $U(1)_X$ is identical to the $I_{3R}$ of $SU(2)_R$ with 
$n_2 = -1/2$. If kinematically alowed, it will be discovered at the LHC.  
Its possible decay into the new particles of Table 2 will then be a good 
signature \cite{cmr07} of this model.

To summarize, it has been shown that the dark scalar doublet model of 
radiative neutrino mass has a supersymmetric realization with the addition 
of a new $U(1)_X$ gauge symmetry.  Here $R$ parity is an automatic 
consequence of the MSSM particle content under $U(1)_X$.  In addition, the 
spontaneous breaking of $U(1)_X$ results in an exactly conserved $Z_2$ 
discrete symmetry which is the basis of the connection between radiative 
neutrino mass and dark matter.  Taken together, this implies at least two 
types of dark matter.  The new particles of this model, especially the 
$U(1)_X$ gauge boson, are accessible to experimental verification at the 
LHC.

This work was supported in part by the U.~S.~Department of Energy under Grant 
No.~DE-FG03-94ER40837.

\bibliographystyle{unsrt}

\end{document}